# Tunable THz generation and enhanced nonlinear effects with active and passive graphene hyperbolic metamaterials

Boyuan Jin[a], Tianjing Guo[a], Liang Zhu[b], Pai-Yen Chen[b], Christos Argyropoulos[*a]

[a]Department of Electrical and Computer Engineering, University of Nebraska-Lincoln, Lincoln, NE, USA 68588; [b]Department of Electrical and Computer Engineering, University of Illinois at Chicago, Chicago, IL, USA 60607

*christos.argyropoulos@unl.edu

## ABSTRACT

The active and nonlinear graphene properties are limited due to weak light-matter interaction between the ultrathin graphene and the incident light. In this work, we present enhanced nonlinear effects at the low terahertz (THz) range by designing a new patterned graphene hyperbolic metamaterial (GHMM). More specifically, it is demonstrated that the third harmonic generation (THG) can be significantly enhanced by the proposed GHMM due to the field enhancement at the resonance as well as the supported slow-light response that fosters strong light–matter interaction.

## 1. INTRODUCTION

Hyperbolic metamaterials (HMM) have recently attracted intense scientific interest in controlling the propagation, scattering, and radiation of light. To date, HMMs have found numerous applications in negative refraction [1, 2], superlensing [3], biosensing [4], perfect absorbers [5], energy harvesting [5, 6], and enhancement of spontaneous emission [7] or Purcell effect [8]. HMM are anisotropic and uniaxial artificial materials which are usually constructed by either metal-dielectric multilayers or metallic nanorod arrays [3]. Metals are crucial to achieve the hyperbolic dispersion, as they provide the desired plasmonic behavior which leads to negative permittivity at optical frequencies.

Graphene is an ultrathin two-dimensional (2D) nanomaterial composed by a single layer of carbon atoms. It exhibits a metallic response at THz and infrared (IR) frequencies. In addition, it has relatively low optical loss, strong optical nonlinearity, enhanced plasmonic resonant response, and tunable complex conductivity via chemical potential (Fermi energy) over a wide spectral range [9-12]. Graphene monolayers can be grown by chemical vapor deposition (CVD) and the graphene's Fermi energy can be tuned via electrostatic gating or chemical doping. Furthermore, population inversion can be achieved in the photoexcited graphene, since the relaxation time of intraband transitions is much shorter than the recombination time of electron-hole pairs [13, 14]. Therefore, for a sufficiently strong photoexcitation, gain can be obtained at THz frequencies due to the cascaded optical-phonon emission and the interband transition around the Dirac point [14-16]. On the other hand, however, graphene cannot strongly interact with the incident excitation wave due to its extreme thinness, i.e., a monolayer of carbon atoms.

In this paper, we present a patterned graphene HMM multilayer metamaterial which can dramatically improve the nonlinear effects at THz frequencies. The proposed metamaterial is built by a periodic array of air-HMM-air waveguides that can support a slow-wave mode in the THz region. Slowing light can promote stronger light–matter interaction and, as a result, boost the nonlinear effects of THz waves. The proposed metamaterial has a resonant response where the electric field can be tightly confined and significantly enhanced. It is demonstrated that the conversion efficiency for the third harmonic generation (THG) (i.e., the ratio of THG power to the excitation power) can be increased by more than three orders of magnitude compared with the unpatterned graphene hyperbolic metamaterials.

## 2. GEOMETRY AND LINEAR PROPERTIES OF THE PATTERNED GHMM

The proposed patterned graphene-based HMM is schematically shown in Fig. 1 which is composed by a periodic array of air-HMM-air waveguides. $N$=10 layers of patterned graphene sheets are stacked and separated by dielectric layers. $P$=1900 nm is the period of the device, $W$=500 nm is the width of the air slots, and $t$=100 nm is the thickness of each dielectric spacer layer. The permittivity of the dielectric spacer is $\varepsilon_d = 2.2$.

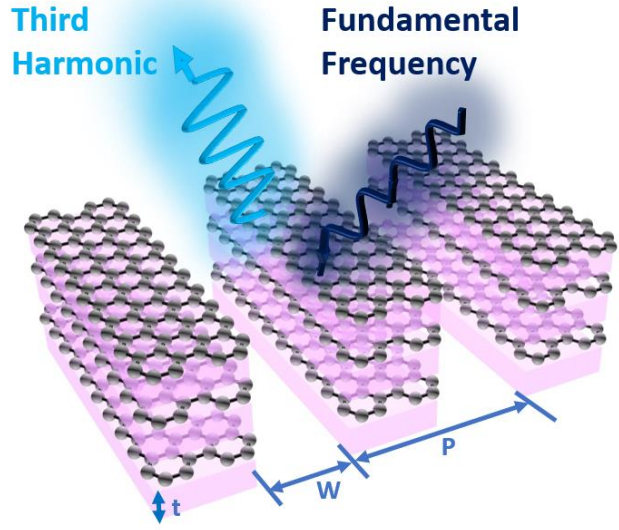

Figure 1. Schematic illustration of the proposed patterned graphene HMM multilayer structure, where $N=10$ patterned graphene sheets are stacked between dielectric layers. $P$ is the period of the metamaterial, $W$ is the width of the air slots, and $t$ is the thickness of each dielectric spacer layer.

The graphene's linear conductivity is given by the formula [16, 17]:

$$\sigma_L = \frac{-jq^2 k_B T}{\pi \hbar^2 (\omega - j\tau^{-1})} \left( \frac{E_F}{k_B T} + 2\ln(e^{-E_F/(k_B T)} + 1) \right) - \frac{jq^2(\omega - j\tau^{-1})}{\pi \hbar^2} \int_0^\infty \frac{f_D(-\varepsilon) - f_D(\varepsilon)}{(\omega - j\tau^{-1})^2 - 4\varepsilon^2/\hbar^2} d\varepsilon, \qquad (1)$$

where $f_D(\varepsilon) = \left[ e^{(\varepsilon - E_F)/(k_B T)} + 1 \right]^{-1}$ is the Fermi-Dirac distribution, and $E_F$ is the Fermi energy or doping level. The calculated frequency dispersion of the real and imaginary conductivity parts for a passive graphene monolayer is shown in Fig. 2. The Fermi level is fixed to 50 meV. The passive linear graphene monolayer exhibits positive real conductivity.

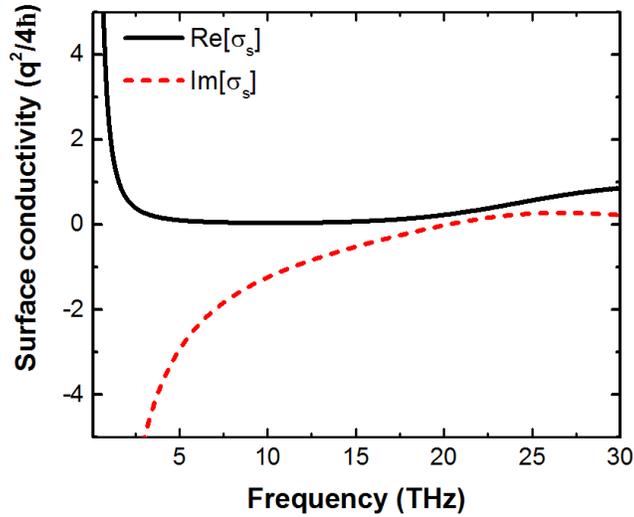

Figure 2. Real and imaginary parts of the linear graphene conductivity as functions of frequency.

The linear spectra of reflectance, transmittance, and absorption of the hyperbolic graphene metamaterial structure shown in Fig. 1 are calculated by COMSOL Multiphysics, a finite element method (FEM)-based commercial software that solves Maxwell's equations. The incident wave is transverse magnetic (TM) polarized. It can be seen in Fig. 3 that the proposed metamaterial is resonant at $f = 5.25$ THz, where the transmittance is close to zero and the reflectance and absorption reach maximum values. At resonance, the electric field is tightly confined along the edges of the graphene patches, as shown in the inset of Fig. 3, with a maximum value of electric field enhancement equal to 37. In addition, it is interesting to note that the thickness of the proposed multilayer graphene HMM is deeply-subwavelength, i.e., $D = 10t = 1\ \mu\text{m} \approx \lambda/60$.

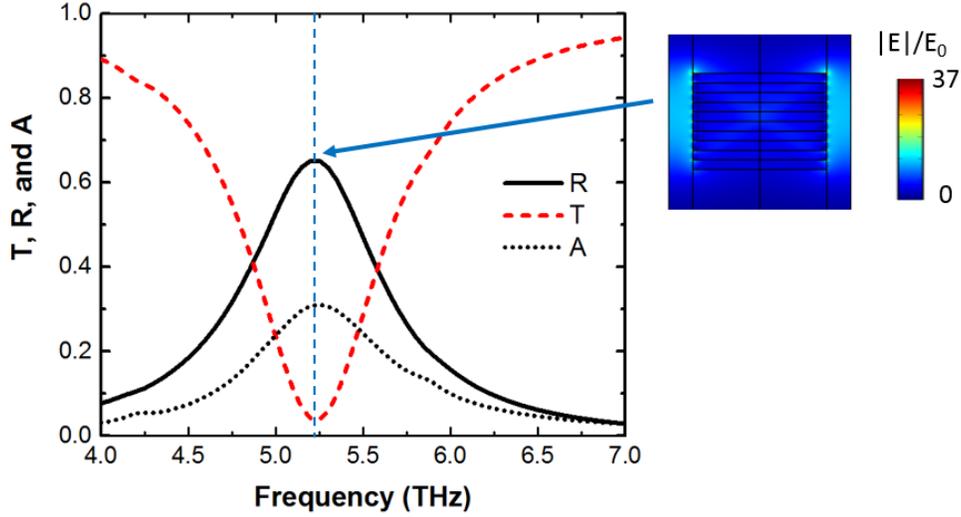

Figure 3. Computed reflectance R, transmittance T, and absorption A as functions of the incident wave frequency. Inset: the computed electric field enhancement $|E|/E_0$ at the reflection peak around 5.25 THz, where E is the local electric field and $E_0$ is the amplitude of the input electric field.

## 3. ENHANCED THG IN THE PROPOSED GHMM

THG is a typical nonlinear wave mixing process, where three photons at the same fundamental frequencies (FF) $f_{FF}$ are annihilated to generate a new photon at the tripled frequency $f_{TH}$, i.e., $f_{TH} = 3f_{FF}$, due to conservation of energy [18]. The conversion efficiency of THG is defined as $CE = P_{r,TH}/P_{i,FF}$, where $P_{i,FF}$ is the incident power at FF. $P_{r,TH}$ is the radiated third harmonic (TH) power outflow calculated by $\int_C \vec{S} \cdot \vec{n}$ with the THG Poynting vector $\vec{S}$ crossing a surrounding to the metamaterial boundary curve C multiplied by the boundary norm vector $\vec{n}$. The nonlinear surface current of graphene can be expressed by [19]:

$$J = \sigma_L(\omega_{TH})E_{FF} + \sigma^{(3)}(\omega_{TH};\omega_{FF},\omega_{FF},\omega_{FF})E_{FF}^3, \quad (2)$$

where $E_{FF}$ is the electric field induced at the FF mode, $\sigma_L$ is graphene's linear conductivity given by Eq. (1). The third-order nonlinear surface conductivity of graphene can be calculated by [20]:

$$\sigma^{(3)} = \frac{i\sigma_0(\hbar v_F q)^2}{48\pi(\hbar\omega_{FF})^4} T(\frac{\hbar\omega_{FF}}{2E_F}), \quad (3)$$

where $\sigma_0 = q^2/4\hbar$, $v_F = 1\times10^6$ m/s is the Fermi velocity, $T(x) = 17G(x) - 64G(2x) + 45G(3x)$ with $G(x) = \ln|(1+x)/(1-x)| + i\pi\theta(|x|-1)$, and $\theta(z)$ is the Heaviside step function. The nonlinear simulations are also conducted by using COMSOL Multiphysics under the assumption of no pump depletion. We ignore the nonlinear response of the dielectric, as it is much weaker than that of the graphene.

We can see from Eq. (3) that the large field enhancement at FF is critical to increase the THG nonlinear effect. As shown in Fig. 3, the electric field is significantly enhanced around the resonant frequency. Therefore, the TH radiation is strongly boosted when the FF is close to 5.25 THz, as shown in Fig. 4. The normalized electric field distribution of the TH mode is shown in the inset of Fig. 4, where the FF equals 5.25 THz and the generated TH frequency is 15.75 THz. Similar to the FF mode, TH field is mainly concentrated at the edges of the graphene patches. In addition, the TH field is also intense at the center of the proposed GHMM.

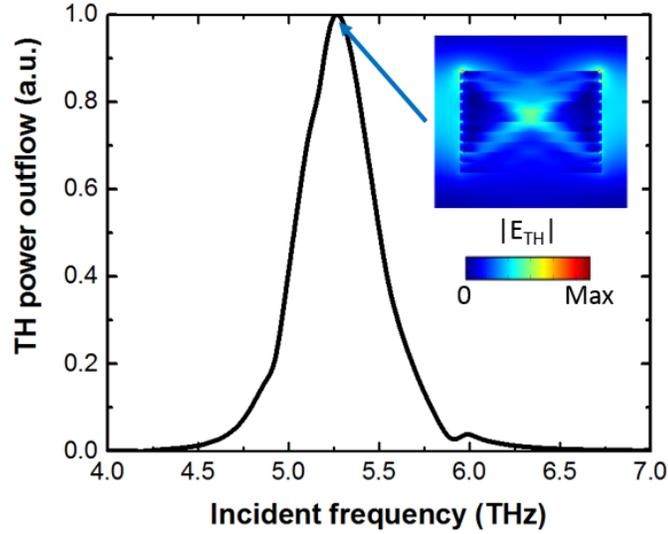

Figure 4. Computed power outflow of the TH radiation as a function of the incident wave frequency. Inset: the computed normalized electric field distributions of the TH wave around the graphene multilayers when the TH radiation is maximum and the corresponding FF is at 5.25 THz.

The THG conversion efficiency (CE) as a function of FF incident intensity $I_0$ is shown in Fig. 5. As expected from Eq. (3), CE is a cubic function of $I_0$. The THG CE can reach to a large value of 2.4% with ultralow input FF intensity $I_0 = 10$ kW/cm². The extremely low required FF intensity is due to the strong nonlinear surface conductivity of graphene and the strong field enhancement of the proposed metamaterial at the resonance. In addition, the proposed graphene HMM can also support a slow-light response [16]. Slowing light can drastically enhance light–matter interactions, such as the THG nonlinear process studied here.

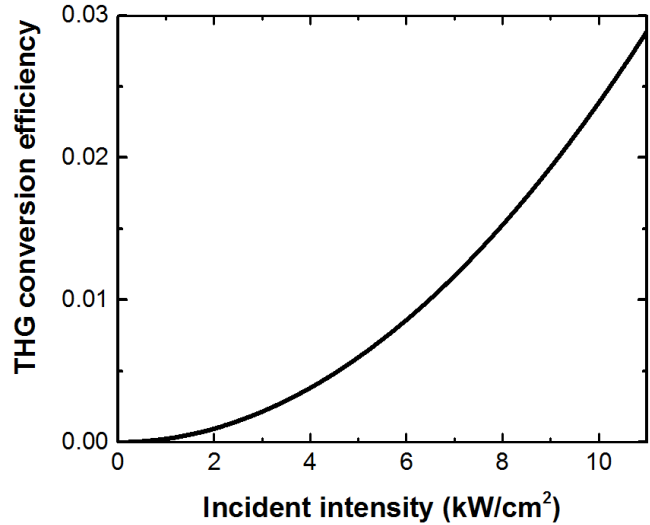

Figure 5. THG conversion efficiency as functions of the incident intensity when the FF frequency is 5.25 THz.

In Fig. 6, we compare the THG CE between the patterned and unpatterned GHMMs. The latter one can be seen as a special case of the patterned GHMM with $W=0$. In both structures, FF equals to 5.25 THz, and we can see that the CE is maximum when the incident FF wave is launched perpendicular to the graphene surface. However, the THG CE is more than three orders of magnitude larger in the proposed patterned GHMM under normal incidence, when compared to the unpatterned GHMM. In addition, the CE is less sensitive to the incident angle in the proposed patterned GHMM.

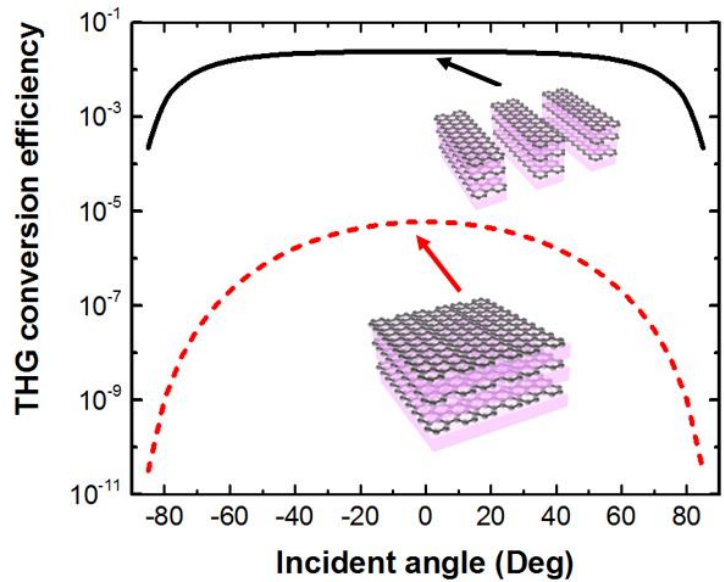

Figure 6. Comparison of the TH conversion efficiency for different incident angles among the proposed patterned graphene metamaterial (black solid line) and the uniform unpatterned graphene HMM multilayer structure where $W=0$ (red dashed line). All the structures are excited with 5.25 THz FF incident waves. The THG conversion efficiency can be increased by more than three orders of magnitude by the proposed metamaterial.

# 4. CONCLUSIONS

To conclude, a GHMM device is proposed based on patterned graphene-dielectric multilayers which can achieve strong THG nonlinear process at THz frequencies. The large field enhancement combined with the strong nonlinear conductivity of graphene, improve the THG CE by several orders of magnitude. The required incident FF intensity to observe significant THG is less than 10 kW/cm$^2$, thus alleviating the typical limitations to achieve high power by THz sources. The proposed GHMM can support a slow-light response, thus enhancing a variety of light–matter interactions, such as the presented THG nonlinear process.